\documentclass[10pt,conference]{IEEEtran}
\usepackage{acronym}

\usepackage{amsthm}

\newcounter{problem}
\newcounter{subproblem}[problem]

\usepackage{amsmath}

\usepackage{etoolbox}

\usepackage{mathtools}
\usepackage[compact]{titlesec}
\titlespacing{\section}{0pt}{0.5ex}{0ex}
\titlespacing{\subsection}{0pt}{0.2ex}{0ex}
\titlespacing{\subsubsection}{0pt}{0ex}{0ex}

\usepackage{color}
\usepackage{xcolor}
\usepackage{colortbl}
\definecolor{purple}{RGB}{139, 0, 139}
\usepackage{manfnt}
\usepackage{url}

\ifCLASSINFOpdf
  \usepackage[pdftex]{graphicx}
   \DeclareGraphicsExtensions{.pdf,.jpg,.png,.tiff}
\else
   \usepackage[dvips]{graphicx}
	 \usepackage{epsfig}
   \usepackage{import}
   \DeclareGraphicsExtensions{.eps,.ps}
\fi
\usepackage{amssymb,amsmath}
\usepackage[mathscr]{euscript}
\usepackage{bbm}
\usepackage{dsfont}

\usepackage{algorithmic}
\usepackage[ruled,vlined,linesnumbered]{algorithm2e}

\usepackage{cite}
\usepackage{subcaption}

\newcommand{\Rmnum}[1]{\uppercase\expandafter{\romannumeral #1}}
\newcommand{\rmnum}[1]{\lowercase\expandafter{\romannumeral #1}}

\newcommand{\field}[1]{\mathbb{#1}}

\newcommand{\emenge}[1]{\mathscr{#1}}

\newcommand{\R}{{\field{R}}}
 
\newcommand{\Ss}{{\emenge{S}}}

\newcommand{\ma}[1]{\boldsymbol{\mathbf{#1}}}
\newcommand{\ve}[1]{\boldsymbol{\mathbf{#1}}}

\newcommand{\va}{\ve{a}}

\newcommand{\vc}{\ve{c}}

\newcommand{\vr}{\ve{r}}
\newcommand{\vp}{\ve{p}}
\newcommand{\vq}{\ve{q}}

\newcommand{\vl}{\ve{l}}

\newcommand{\vm}{\ve{m}}

\newcommand{\vo}{\ve{o}}

\newcommand{\mI}{\ma{I}}

\newcommand{\mR}{\ma{R}}

\newcommand{\mK}{\ma{K}}

\newcommand{\mP}{\ma{P}}

\usepackage[utf8]{inputenc}

\usepackage{amsthm}
\usepackage{amssymb,amsmath}
\usepackage[mathscr]{euscript}

\newcommand{\cosl}[1]{}
\newcommand{\resl}[1]{}

\usepackage[draft,footnote,nomargin]{fixme}
\newcommand{\fnql}[1]{}
\newcommand{\fnsv}[1]{}

\begin{document}
%
\title{HARU: Haptic Augmented Reality-Assisted User-Centric Industrial Network Planning\vspace{-0.3in}}
\author{
\IEEEauthorblockN{Qi Liao\IEEEauthorrefmark{1}, 
Nikolaj Marchenko\IEEEauthorrefmark{2}, 
Tianlun Hu\IEEEauthorrefmark{1},
Peter Kulics\IEEEauthorrefmark{1},
Lutz Ewe\IEEEauthorrefmark{1}}
\IEEEauthorblockA{ 
	\IEEEauthorrefmark{1}Nokia Bell Labs, Stuttgart, Germany\\
	\IEEEauthorrefmark{2}Corporate Research and Advanced Engineering, Robert Bosch GmbH, Renningen, Germany\\
E-Mails: \{qi.liao, peter.kulics,  lutz.ewe\}@nokia-bell-labs.com, \\
	   Nikolaj.Marchenko@de.bosch.com,
	   tianlun.hu@nokia.com \vspace{0.1in}}
}


%


\maketitle
\begin{abstract}
To support Industry 4.0 applications with haptics and human-machine interaction, \ac{6G} requires a new framework that is fully autonomous, visual, and interactive. In this paper, we provide an end-to-end solution, HARU, for private network planning services, especially industrial networks. The solution consists of the following functions: collecting visual and sensory data from the user device, reconstructing 3D radio propagation environment and conducting network planning on a server, and visualizing network performance with \ac{AR} on the user device with enabled haptic feedback. 
The functions are empowered by three key technical components: 1) vision- and sensor fusion-based 3D environment reconstruction, 2) ray tracing-based radio map generation and network planning, and 3) \ac{AR}-assisted network visualization enabled by real-time camera relocalization. We conducted the proof-of-concept in a Bosch plant in Germany and showed good network coverage of the optimized antenna location, as well as high accuracy in both environment reconstruction and camera relocalization. We also achieved real-time \ac{AR}-supported network monitoring with an end-to-end latency of about $32$ ms per frame. 
\end{abstract}

\IEEEpeerreviewmaketitle

\section{Introduction}\label{sec:intro}
Impacted by the ever-increasing global spending on smart manufacturing, the total addressable market for private networks is forecast to increase from $\$3.7$ billion in 2021 to over $\$109.4$ billion in 2030, according to a report by ABI Research \cite{ABIrep}. However, the state-of-the-art network planning services remain old-fashioned. For example, the network planning service providers rely on either user-uploaded floor plan or on-site measured site survey. Not only are such services costly in both time and human resources,  but they also provide a limited user experience. Moreover, the modern flexible and modular production systems require industrial network planning solutions to be quickly adapted to new network environments triggered by dynamically changing batches of individual products.

Extended reality and digital twin can be promising technologies to facilitate autonomous and interactive network services in the \ac{6G} era \cite{viswanathan2020communications,nguyen2021digital}. However, most of the works provided perspective on the challenges they raise to future network systems, e.g.,  as emerging 5G use cases \cite{mahmoud20216g}, while few of the works considered them as a means to help build truly user-centric, interactive cyber-physical network systems \cite{koutitas2020situ}.  A few works proposed conceptual \ac{AR}-based frameworks for various small-scale network solutions. For example, WiART \cite{nguyen2016wiart} takes inputs from beam-steerable reconfigurable antennas and enables users to select desired antenna radiation patterns in the mobile app and observe their effects on link performance. However, the \ac{AR} visualization is limited to the antennas and their radiation patterns within a short distance. In \cite{koutitas2020situ}, the authors proposed to scan an indoor apartment environment to a vector facet model and use it for wireless signal propagation prediction with ray tracing. Then, an \ac{AR} application is applied to visualize the prediction in the form of holograms with an \ac{AR} headset. However, the performance is limited by the lack of information about the obstacles for radio propagation. Also, the solution requires licensed \ac{AR} applications and expensive commercial \ac{AR} headsets.
\begin{figure}[t]
    \centering
    \includegraphics[width=.45\textwidth]{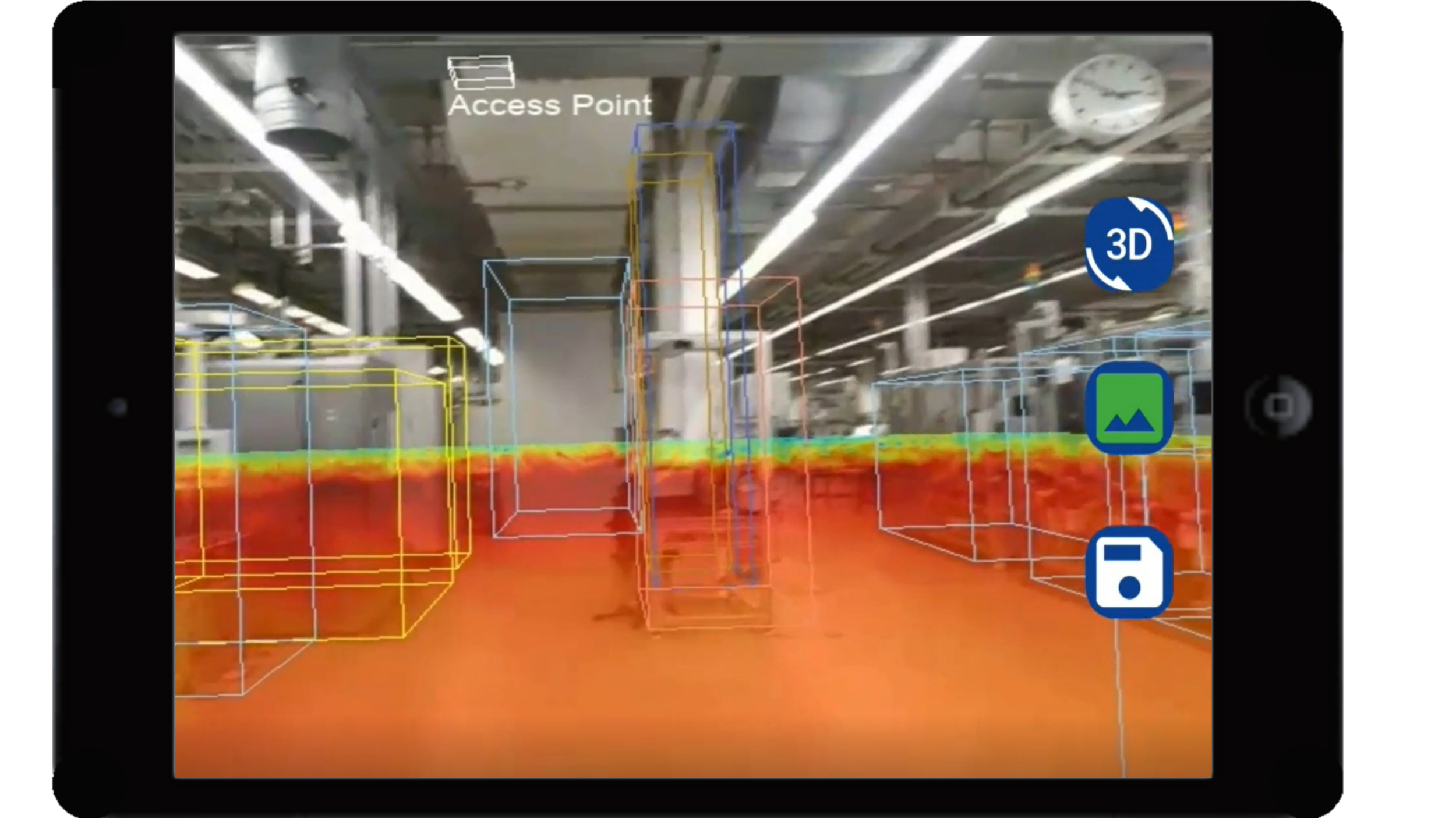}
    \caption{Haptic AR-assisted network planning app}
    \label{fig:AP-APP}
    \vspace{-5.5ex}
\end{figure}
\begin{figure*}[t]
    \centering
    \includegraphics[width=.75\textwidth]{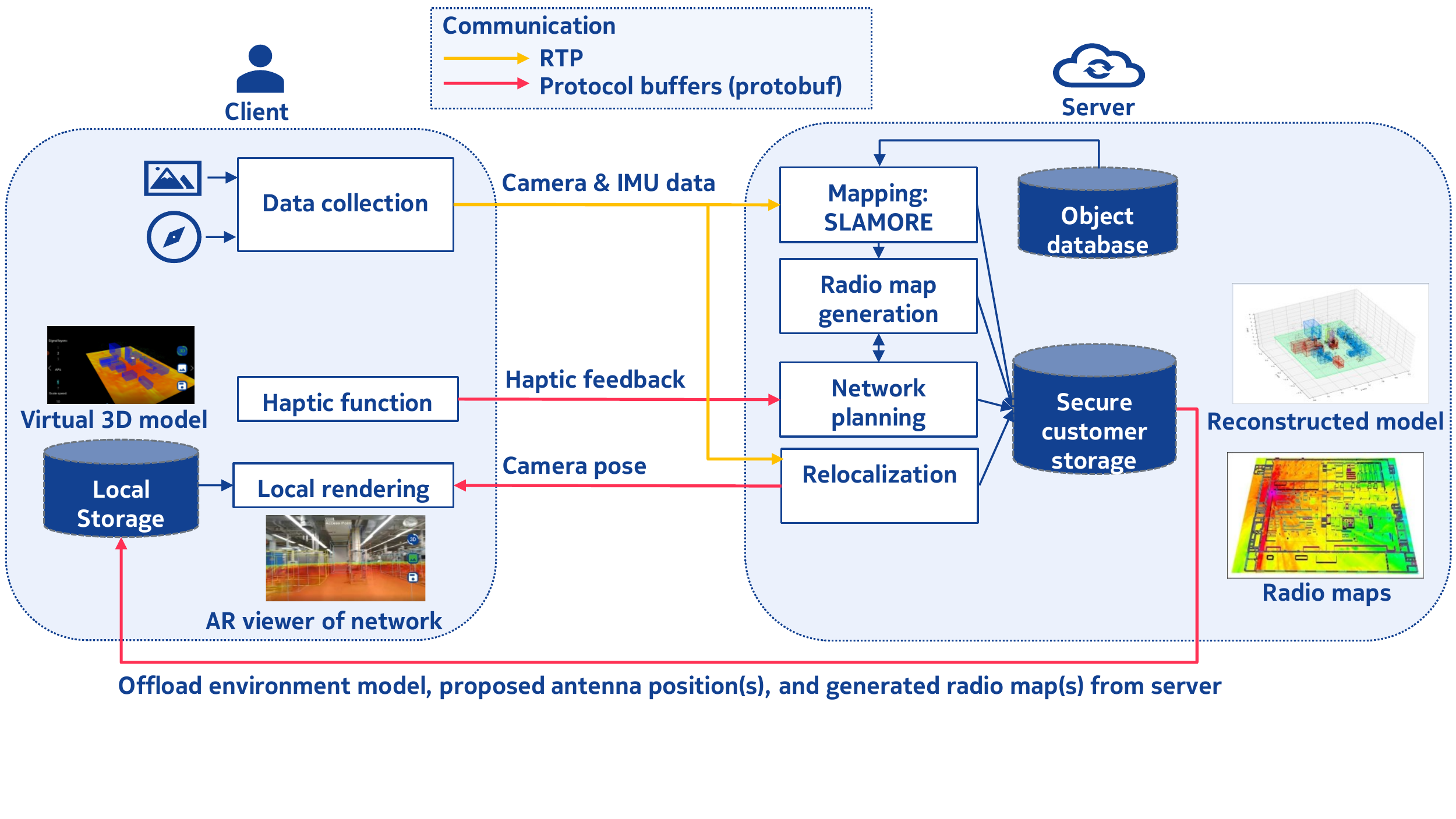}
    \vspace{-8ex}
    \caption{Architecture of the end-to-end \ac{AR}-empowered network planning solution}
    \vspace{-5ex}
    \label{fig:E2E_Architecture}
\end{figure*}
To provide autonomous, visual, and interactive service for private network planning with low-cost mobile devices, in \cite{liao2018method} and \cite{liao2019haptic}, we proposed HARU, a haptic augmented reality-assisted user-centric network planning solution,  enabled by \emph{vision- and sensor fusion-based 3D radio environment reconstruction} and \emph{\ac{AR}-assisted network visualization}. The goal is to make the invisible radio network visible and provide truly user-centric, gamified, haptic interface to improve user experience. The service platform receives video/image sequence and \ac{IMU} data from the user device via a mobile \ac{UI}, reconstructs the 3D radio environment based on the received data, approximates the radio propagation model, and demonstrates the optimized network planning solution based on the user's haptic feedback on \ac{UI} with \ac{AR}, as shown in Fig. \ref{fig:AP-APP}. 
A prerequisite for enabling \ac{AR} is to estimate \emph{6D camera pose} (including 3D position and 3D orientation) in real time, known as the \emph{camera relocalization} problem. With the estimated camera pose, we can project the virtual objects from the real-world coordinate system back to the camera coordinate system.
In our previous works \cite{Liao20,HuCamLoc21}, we have provided technical solutions and algorithms for individual functions, \emph{3D radio environment reconstruction} and \emph{real-time camera relocalization}, respectively. In this paper, we focus on the system design of the end-to-end solution and its application to a real industrial environment. Note that in the previous works, we conducted the experiments in a small-scale office environment only, while in this paper we demonstrate the most recent results of the \ac{PoC} in a Bosch plant in Germany. 
 %
%
To the best of the authors' knowledge, we provide the first end-to-end interactive \ac{AR}-assisted solution to industrial network planning.  Our contributions are summarized as follows.
\vspace{-1ex}
\begin{itemize}
\itemsep0.3em 
\item On the client side, we developed an Android app, which sends visual and sensory data from the mobile device to a server, receives model and localization data from the server, and performs local rendering and haptic function for the interactive interface. 
\item On the server side, we developed and implemented the following functions: 1) a {\bf 3D environment reconstruction} algorithm called  \ac{SLAMORE} \cite{Liao20}, which can detect, track, localize, and reconstruct the major obstacles for electromagnetic waves, 2) a {\bf network planning policy} assisted by ray-tracing application, and 3) a {\bf real-time camera relocalization} algorithm \cite{HuCamLoc21} based on deep learning and sensor fusion to enable the \ac{AR} features.  
\item We demonstrated the \ac{PoC} of the end-to-end solution with a standard Android mobile device and the edge server in the {\bf real industrial environment} -- a Bosch plant in Germany, and achieved an end-to-end delay (including data compression, communication, and computation) of about $32$ ms per frame.
\end{itemize}
%
\section{System Architecture}\label{sec:architect}
The proposed haptic \ac{AR}-assisted network planning service platform includes three main techniques: 1) mapping and 3D industrial environment reconstruction, 2) ray tracing-based radio map generation and network planning, and 3) real-time camera relocalization for \ac{AR} features. Because the remote service aims to provide real-time \ac{AR} experience for monitoring and interacting with the network, we need to allocate the different functions in either the mobile device or the server based on their computational, transmission, and latency constraints. After extensive experiments, we design the system architecture of the end-to-end solution as shown in Fig. \ref{fig:E2E_Architecture}. We decide to load the reconstructed 3D model and radio maps from the server to the device, such that the \ac{AR} rendering function can be called locally. In general, local rendering function responds much faster to user's inputs for interaction, compared to remote rendering. Moreover, because our \ac{SLAMORE} algorithm only extracts and reconstructs objects relevant to radio propagation, the size of data is small (less than $5$ MB) to be transferred and locally stored. On the other hand, camera relocalization is performed on the server because the deep learning-based model is relatively large and real-time local inference is computationally inefficient. 

In the first step ({\bf mapping}), the user runs our app on a mobile device and the app sends the captured visual and sensory data to the server.  The server receives the data and calls the mapping function -- our developed \ac{SLAMORE} algorithm \cite{Liao20} -- to detect, track, localize, and reconstruct the major obstacles for electromagnetic waves. In this way, the server reconstructs an \lq\lq extracted\rq\rq \ version of the environment customized for radio propagation. Such a sparse reconstruction of the environment significantly eases the computation of ray tracing for further steps of radio map generation and network planning.  

In the second step ({\bf interactive network planning}),  with the haptic function, the user can specify a preferred area (e.g., where a power supply is available) for \ac{AP} deployment and send the information to the server. The server generates radio maps with ray tracing based on the reconstructed 3D environment, and performs network planning policy constrained by the user-specified area. Then, the server sends the extracted environment model, the proposed \ac{AP} deployment position(s), and the corresponding generated radio map(s) to the user. The data is saved in the local storage and can be retrieved locally to enable haptic interaction with the virtual 3D model of the network environment. 

In the final step ({\bf  \ac{AR}-assisted network monitoring}),  the server trains a multi-input \ac{DNN} for camera relocalization \cite{HuCamLoc21}, based on the previously collected data and the generated environment in the mapping step. The user sends the visual and sensory data to the server in real time, and the relocalization model takes it as input and estimates the 6D camera pose as output. The server then sends the estimated camera pose to the user device, and the app computes rendering locally and projects the augmented radio map on the captured camera view.

\section{End-to-End Solution}\label{sec:solution}
In this section, we describe the following technical components of the end-to-end solution: 1) data collection and preprocessing, 2) mapping and 3D radio environment reconstruction, 3) 3D radio map generation and network planning, and 4) real-time camera relocalization. 
\subsection{Data Collection and Preprocessing}\label{subsec:DataCollect}
The app running on mobile device captures in real time the RGB camera stream and the motion sensor measurements, including \emph{timestamp}, \emph{3D angular rate}, \emph{3D linear acceleration}, and \emph{4D quaternion} from the device, and sends them to the server via \ac{RTP} (details of the implementation will be given in Section \ref{ssec:commProto}). These measurements are used for both 3D environment reconstruction and camera relocalization algorithms. 

On the server, to facilitate \ac{SLAMORE} and reconstruct an obstacle-aware radio propagation environment, a labelled dataset of finite objects in the industrial environment needs to be pre-collected. One may argue that in practice the pretrained set of object classes can be too large for recognizing all obstacles in the new environments. However, because we target private networks, we can train multiple sets of classes for different types of customers. For example, automotive manufacturing plants from various manufacturers possibly have deployed similar types of equipment, thus we can create a large dataset for this specific type of customers.
\begin{figure}[t]
    \centering
    \includegraphics[width=.49\textwidth]{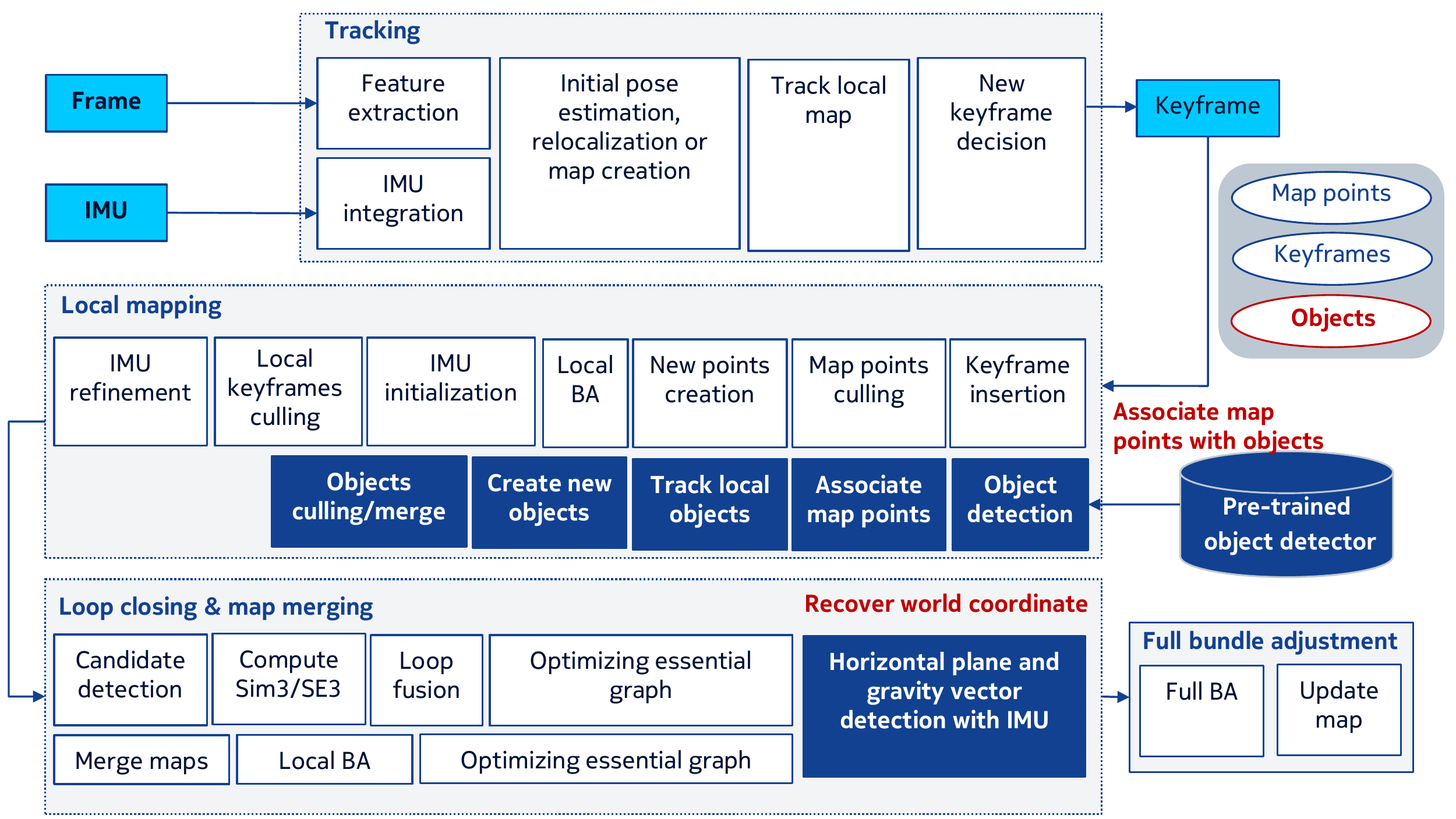}
    \caption{\ac{SLAMORE} architecture, where Sim3/SE3 stands for 3D affine transform and BA means the bundle adjustment. The white blocks are the original functions from ORB-SLAM3 \cite{campos2021orb}, while the blue blocks are our developed and embedded functions for 3D object reconstruction and plane detection.}
    \label{fig:SLAMORE}
    \vspace{-5ex}
\end{figure}
\subsection{Mapping and 3D Radio Environment Reconstruction}\label{subsec:EnvRecon}
 Most of the mapping and environment reconstruction methods, such as 3D scanning or \ac{SLAM} algorithms, provide complex reconstructed models by creating ultra-dense 3D point cloud (a collection of points defined by a given coordinate system) and recovering smooth nonlinear surfaces \cite{berger2017survey}. However, for radio propagation environment reconstruction we cannot afford such complexity: modeling radio propagation maps, e.g., by using ray tracing \cite{yun2015ray}, can be computationally extremely expensive if it is based on a complex environment. 
 Thus, to customize an environment reconstructor for 3D radio propagation modeling, we target the following problem: 

\emph{How to efficiently reconstruct an \lq\lq extracted\rq\rq \ industrial radio propagation environment that achieves good accuracy yet eases the task of 3D radio propagation modeling?} 

To this end, we propose a feature- and object-based monocular \ac{SLAM} algorithm called \ac{SLAMORE} \cite{Liao20}, which can efficiently detect, track, localize, and reconstruct the major obstacles for electromagnetic waves. It extracts the industrial network environment composed of reconstructed obstacles for radio propagation and detected space boundaries. \ac{SLAMORE} is built on the basis of a state-of-the-art visual-inertial \ac{SLAM} algorithm ORB-SLAM3 \cite{campos2021orb}. However, ORB-SLAM3  reconstructs the environment with sparse 3D point cloud and the output point cloud does not provide the information needed for radio propagation modeling, such as real-world coordinates and scale, object segmentation, and surface material. To overcome this challenge, we propose \ac{SLAMORE} with the following new features, as shown in Fig. \ref{fig:SLAMORE}:
1) We train a {\it customized convolutional object detector} for a defined set of objects in industrial radio propagation environment. The label includes the class of the object, 3D shape and size, and surface material. 
2)  A series of new functions are embedded in the local mapping thread for 3D object reconstruction and called for every keyframe, including \emph{object detection}, \emph{associating map points to objects}, \emph{object tracking},  \emph{new object creation}, and \emph{object culling and merging}.
3) We solve the major problem in ORB-SLAM3 when using low-cost and low-frequency \ac{IMU} sensors from mobile device -- {\it recovery of the real-world coordinates and scale} -- by using the side information contained in the class labels of the recognized objects and by filtering the \ac{IMU} measurements.

With these new features, we can extract sufficient but not overwhelming information about the environment, to be further used as the input for the efficient radio propagation modeling. Due to the limited space, we refer the interested readers to our previous work \cite{Liao20} for the technical details.
\subsection{3D Radio Map Generation and Network Planning}\label{subsec:MapGenNetPlan}
Because ray tracing for radio propagation modeling is a well-studied topic, we can use existing algorithms or \acp{SDK} to compute the radio map such as WinProp \cite{hoppe2017wave} and WISE \cite{507047}. Also, to further reduce complexity, various shooting-and-bouncing ray tracing algorithms \cite{kasdorf2021advancing} can be applied. Moreover, because \ac{SLAMORE} is customized for reconstructing radio propagation environment, it allows more efficient computation of ray tracing, comparing to the complex 3D scan composed of huge nonlinear triangular meshes.  With ray tracing, we can obtain the \ac{RSRP} values of any point in the 3D space. For efficient network planning, we can discretize the bounded 3D space $\Ss:=\left[x^{(\min)}, x^{(\max)}\right]\times \left[y^{(\min)}, y^{(\max)}\right]\times \left[z^{(\min)}, z^{(\max)}\right]$ into equally-spaced nonoverlapping voxels. Thus, the 3D \ac{RSRP} map can be computed for each antenna position $\va\in\Ss$, denoted by $\mP(\va):= [p_{i,j,k}(\va)]$, where $p_{i,j,k}(\va)$ denotes the \ac{RSRP} at the voxel with indices $(i,j,k)$
in $x, y, z$ dimensions, respectively. 

The network planning aims to provide good coverage of the factory by optimizing the position of \ac{AP} $\va\in\Ss$, especially where the machines/devices equipped with transceivers are located. We define the utility function as the weighted sum of \ac{RSRP} over all voxels, while assigning higher weights to the following types of voxels: 1) those associated to the machine objects which generate network traffic, and 2) those associated to the defined trajectory of mobile objects such as \acp{AGV}. The optimization problem is then defined as
\begin{equation}
    \max_{\va\in\Ss}U(\va), \mbox{ s.t. } U(\va)=\sum_{i,j,k}w_{i,j,k} p_{i,j,k}(\va).
    \label{eqn:planningprob}
\end{equation} 
Because of the interactive interface and the haptic functions, the users can specify their preferred deployment areas, e.g., a limited area on the ceiling, denoted by $\Ss'$. Thus, the searching space $\Ss$ is further reduced to $\Ss'$ and we can replace $\Ss$ with $\Ss'$ in \eqref{eqn:planningprob}. We consider the gradient-based search over $\Ss'$. For efficient searching, we define the initial point as the geometric center $\vc$ of all detected machines/devices. Then, we generate $N$ searching instances, each starts from a distinct initial position sampled near $\vc$. We perform gradient-based search for each of the instances, and choose the one with the maximum converged utility among all instances. Note that the gradient-based search is an iterative process, i.e., we need to compute the radio map for each iterative \ac{AP} position, thus a fast yet accurate ray tracing computation is crucial to reduce the complexity.  \ac{SLAMORE} well fulfills the task, because it extracts sufficient but not overwhelming information about the environment, which further enables faster ray tracing computation.     
\subsection{Real-Time Camera Relocalization to Enable AR}\label{subsec:CamLoc}
\begin{figure}[t]
    \centering
    \includegraphics[width=.45\textwidth]{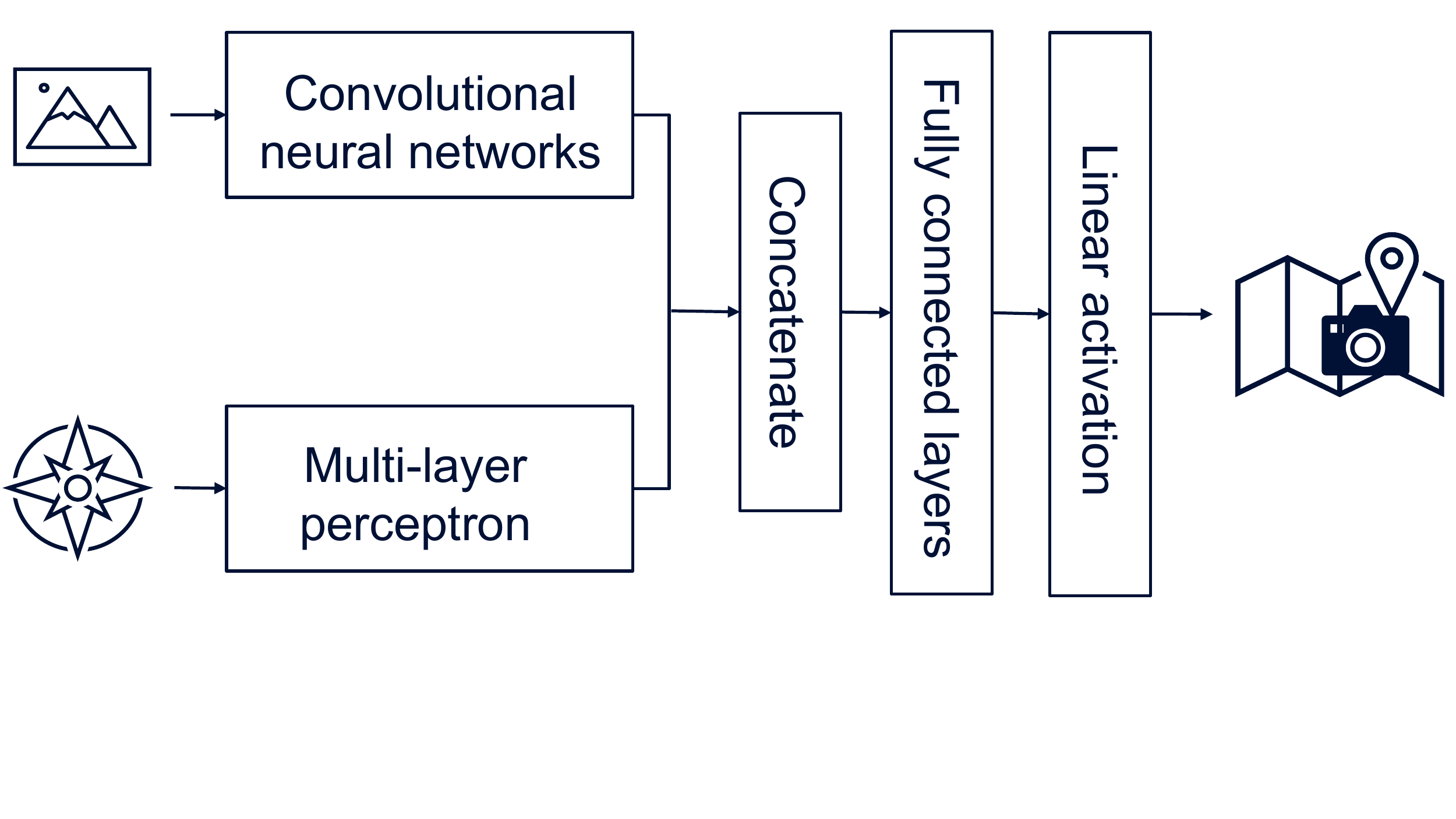}
    \vspace{-7ex}
    \caption{Multi-input \ac{DNN} for real-time camera relocalization}
    \label{fig:multiInputDNN}
    \vspace{-4ex}
\end{figure}
To enable the \ac{AR} features, e.g., overlaying the virtual radio map on the 2D images captured by camera, we need to estimate the 6 \ac{DoF} camera pose $\vp := [\vl, \vo]$ consisting of camera location $\vl$ and orientation $\vo$ in real time. Therefore, we are interested in solving the following problem:

\emph{At each time frame $t$, given camera captured image array $\mI(t)\in\R^{w \times h \times 3}$, where $w$ and $h$ are the image width and height in pixel respectively, and the extracted feature vector from motion sensors $\vm(t)\in\R^n$, what is the corresponding camera pose $\vp(t) := [\vl(t), \vo(t)]$  composed of camera location $\vl:=[x, y, z]$ and orientation $\vo$?} 

Note that the orientation $\vo$ can be written in the form of 3D rotation vector $\vr:=[r_x, r_y, r_z]$ or its equivalent 4D quaternion $\vq := [q_x, q_y, q_z, q_w]$. The quaternion $\vq$ needs to be normalized to unit length, thus the camera pose is still $6$-\ac{DoF} when $\vo=\vq$. 

To solve the problem, we propose a multi-input \ac{DNN} including both image and sensor data in the inputs. The architecture of the \ac{DNN} is shown in Fig. \ref{fig:multiInputDNN}. The input image is resized to $(224, 224, 3)$ and fed into
a pretrained \ac{CNN} backbone for object detection, such as MobileNetV2 \cite{sandler2018mobilenetv2}. To accelerate the learning, we load the weights of the pretrained \ac{CNN} object detector for \ac{SLAMORE} as initial weights, since it is customized for the predefined object classes in the interested industrial environment. The sensor input is then passed through a \ac{MLP} composed of three dense (fully connected) layers with ReLU activation. The output layers of the CNN and MLP are concatenated and followed with $5$ more dense layers with ReLU activation. Finally, the output layer with linear activation provides the camera pose estimation. The training data can be obtained from the data previously collected for \ac{SLAMORE}, including the video sequence, the \ac{IMU} data, and the corresponding camera pose in the world coordinate system. Due to the limited space, we omit the details here but refer the interested readers to our previous work \cite{HuCamLoc21}.

After deriving the estimated camera pose in the world coordinate system, denoted by location $\vl_w\in \R^3$ and rotation $\mR_{cw}\in\R^{3\times 3}$ (rotation matrix from the camera coordinate system to the world coordinate system, converted from orientation $\vo$), we can project any 3D point of radio map $\vp_w$ in the world coordinate system into the pixel system $(u,v)$ as below:
\begin{align}
    \begin{bmatrix}
u'\\
v' \\
\alpha
\end{bmatrix}
& = \mK \mR_{cw}^T \left[\vp_w - \vl_w\right],\\
u & = u'/\alpha \mbox{ and } v = v'/\alpha,
\end{align}
where $\mR_{cw}^T=\mR_{wc}$ indicates the rotation matrix from the world to the camera coordinate system,  $\mK\in\R^{3\times 3}$ is the camera intrinsic matrix, and $\alpha\neq 0$ is the scaling factor learned by \ac{SLAMORE}.

\section{Proof-of-Concept}\label{sec:PoC}
We conducted the experiment in a Bosch plant in Germany, and selected an area of $15\text{m}\times10\text{m}$. The applied antenna model is Nokia FW2HC integrated omni antenna with antenna gain of $4.7$ dBi. We use Nokia $6$ as the mobile device and the edge server is equipped with $4$ Nvidia Tesla $K80$ GPUs for \ac{SLAMORE} and \ac{DNN} training and inference.
\subsubsection{Data Collection and Communication Protocols}\label{ssec:commProto}
Our developed Android app captures RGB camera stream with a resolution of $480$p and the motion sensor measurements including \emph{timestamp}, \emph{3D angular rate}, \emph{3D linear acceleration} (from accelerometer), and \emph{4D quaternion} (from Android attitude composite sensors which derive the rotation vector from the physical sensors accelerometer, gyroscope, and magnetometer). Note that many mobile devices are equipped with low-cost magnetometers and the reported quaternion can be very noisy. A low-pass filter is applied to the sensor output to reduce noise artifacts and smooth the signal reading. We use \ac{RTP} to carry the video stream compressed by codec. The sensory data is also included in the \ac{RTP} packets along with the video stream since the \ac{RTP} specification allows for a custom extension to the protocol.
\subsubsection{3D Radio Environment Reconstruction}
\begin{figure}[t]
    \centering
    \includegraphics[width=.35\textwidth]{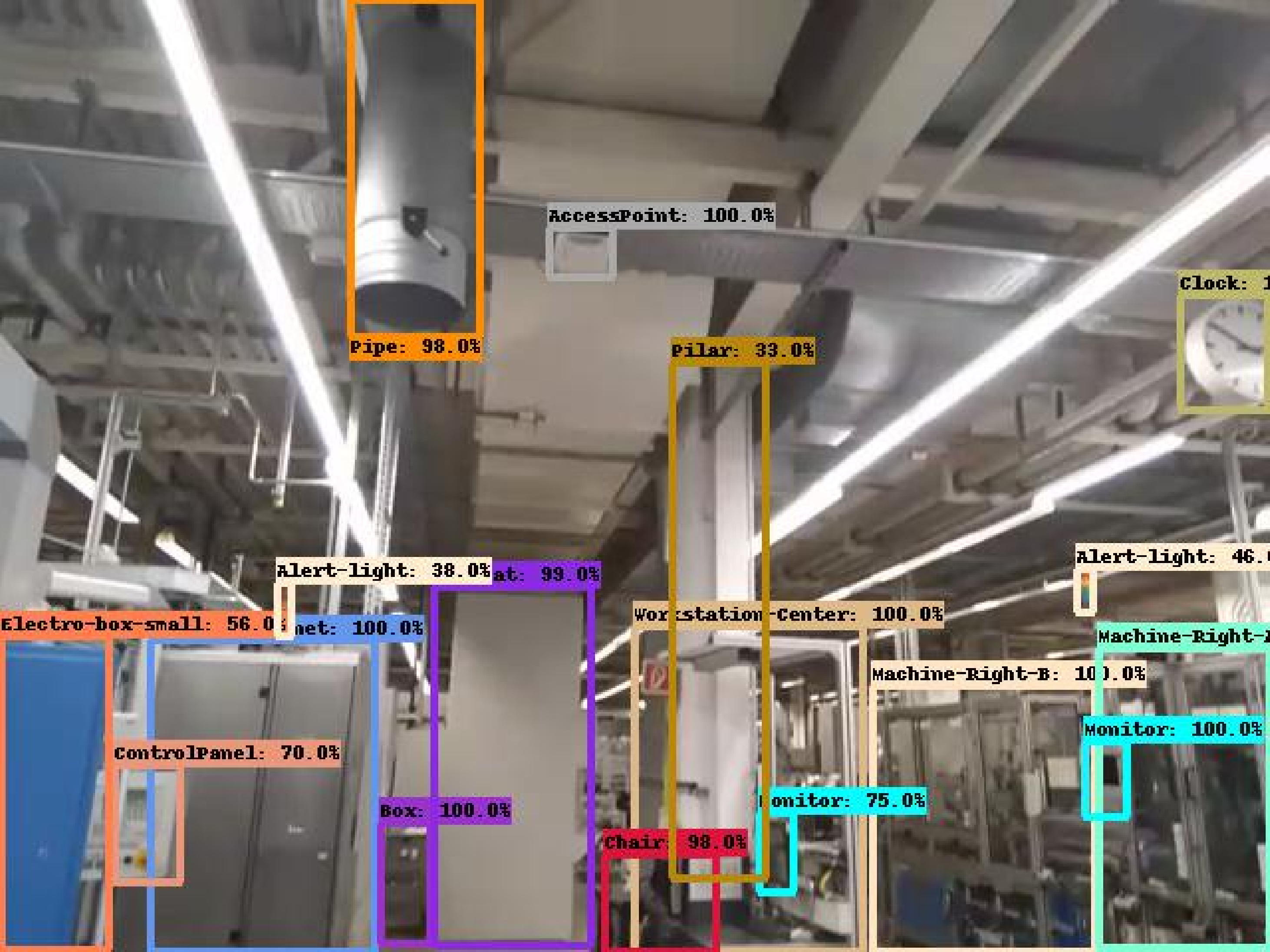}
    \caption{Object detection customized to industrial environment}
    \label{fig:objectDetect}
    \vspace{-3ex}
\end{figure}

To test \ac{SLAMORE}, we collected $5$ video sequences, which provided $80596$ frames in total. For training the object detector, we manually labelled only $300$ keyframes with $25$ classes of machines and objects in the industrial environment. The object detector is initialized with the SSD MobileNetV2 \cite{sandler2018mobilenetv2} backbone model from Tensorflow 2 Object Detection API and finetuned with our labelled data collected from the industrial environment. Although the number of labelled training data is limited, we are able to achieve $87.2\%$ mean average precision. An example of the detected objects in a frame is given in Fig. \ref{fig:objectDetect}. 

Then, for every keyframe, we project the detected map points in the 3D space back into the 2D camera view and adaptively associate them to the bounding boxes detected by the object detector. By clustering the 3D map points associated to the objects, and, with the shape information contained in the object's label, we can estimate the position of the objects. We evaluate the performance with the average object-based \ac{RMSE} defined as
$(1/K)\cdot\sum_{k=1}^K \sqrt{\sum_{i\in\{x,y,z\}} (p_i - \hat{p}_i)^2/3}$, where $K$  is the number of detected objects, $p_i$ and $\hat{p}_i$ are the ground-truth and the estimated object's position in $i\in\{x, y, z\}$ axis. We achieve the averaged \ac{RMSE} of $0.2564$ m over all detected objects within the selected area of 15m × 10m.   

\begin{figure}[t]
    \centering
    \includegraphics[width=.48\textwidth]{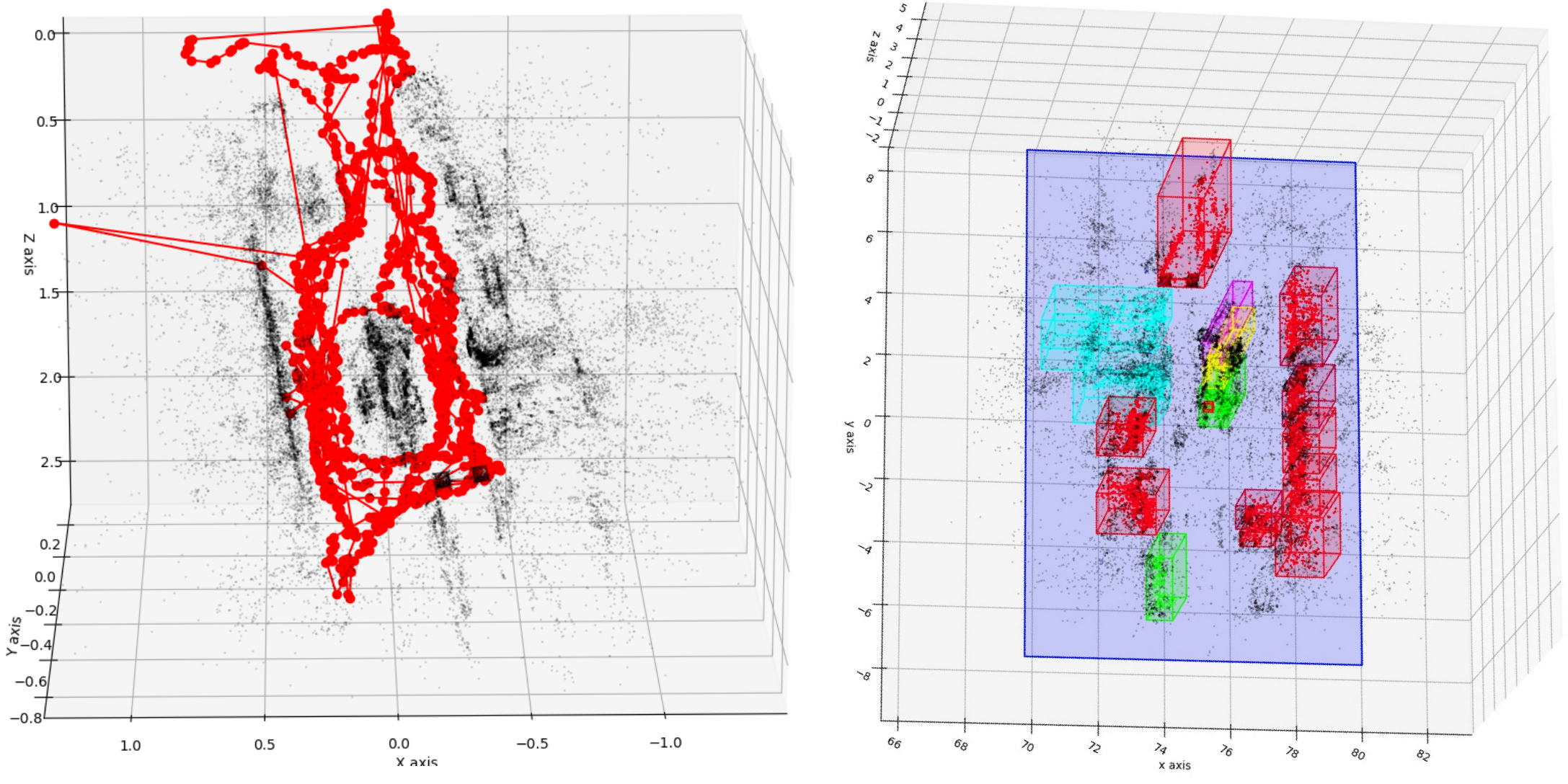}
    \caption{[Left] Unaligned coordinates with detected map points with ORB-SLAM3. [Right] Reconstructed environment with the recognized 3D obstacles using \ac{SLAMORE}.}
    \label{fig:model_map}
    \vspace{-4ex}
\end{figure}

In Fig. \ref{fig:model_map} we compare the output of our proposed \ac{SLAMORE} with the state-of-the-art output of the mapping computed from ORB-SLAM3 \cite{campos2021orb}. On the left side of the figure we show the raw outputs of ORB-SLAM3, including the map points as the black points and the estimated camera positions as the red trajectory, with inaccurate alignment with the world coordinates and without object segmentation. The poor alignment is caused by the noisy measurements from the low-cost \ac{IMU} sensors in the mobile device. On the right side we show the output of \ac{SLAMORE}, providing the reconstructed environment and obstacles for radio propagation environment, and the map points associated to distinct obstacles.
%
%


\subsubsection{3D Radio Map Generalization and Network Planning}
\begin{figure}[t]
    \centering
    \includegraphics[width=.35\textwidth]{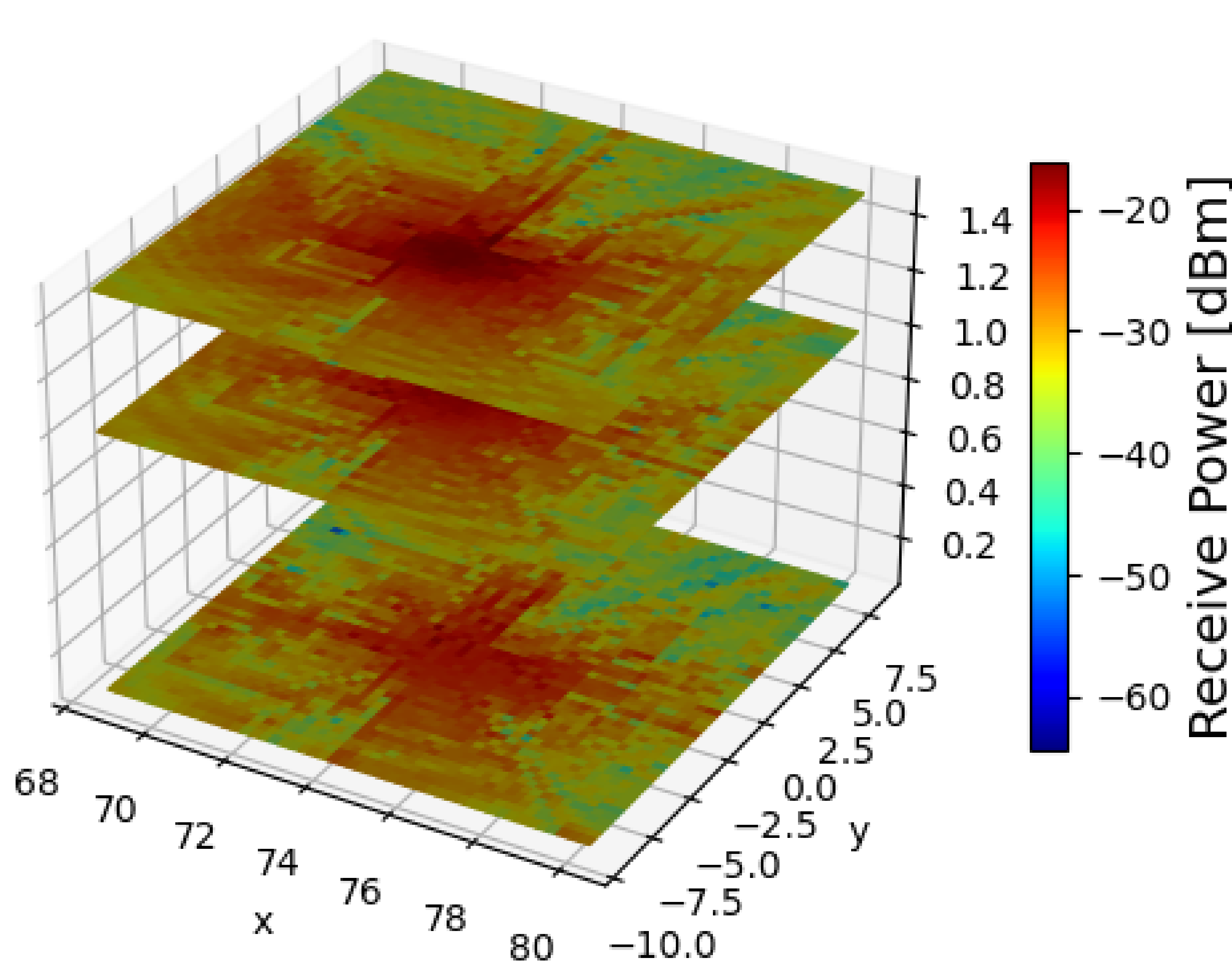}
    \caption{Generated radio map of the optimized \ac{AP} position}
    \label{fig:generatedMap}
    \vspace{-3ex}
\end{figure}

\begin{figure}[t]
    \centering
    \includegraphics[width=.48\textwidth]{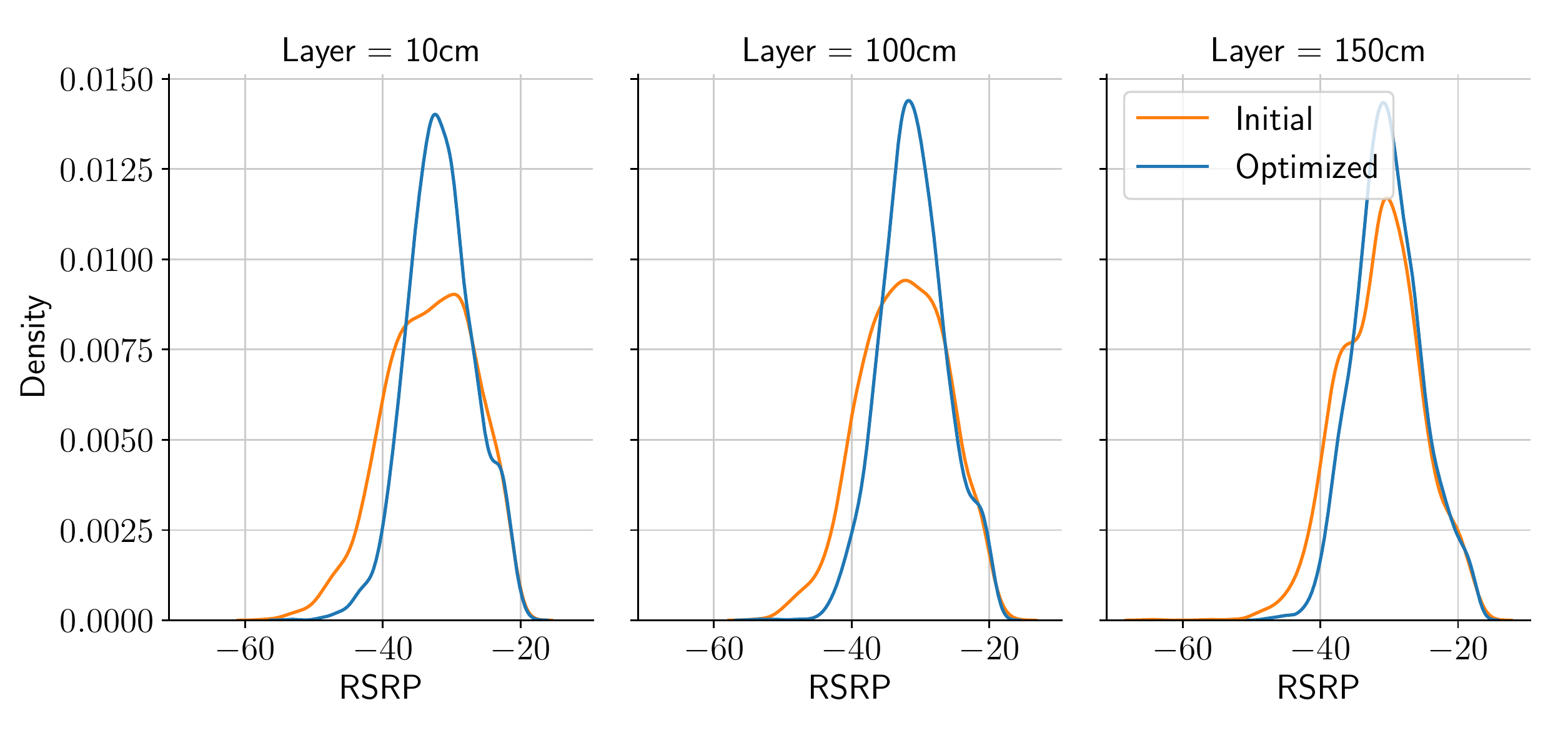}
    \caption{Comparison of the \ac{RSRP} distribution between initial and optimized antenna positions at different heights.}
    \label{fig:rsrp_dist}
    \vspace{-3ex}
\end{figure}

In this experiment, we call the functions from WinProp \cite{hoppe2017wave} to compute the radio map with ray tracing technique. The heuristic searching algorithm described in Section \ref{subsec:MapGenNetPlan} converges between $20$ and $40$ steps for each randomly initialized instance, which, in total costs less than $1$ hour to perform the network planning task for the 3D industrial space. The generated radio maps of the optimized antenna position at different heights $10$ cm, $100$ cm, and $150$ cm are shown in Fig. \ref{fig:generatedMap}. In Fig. \ref{fig:rsrp_dist} we show that the optimized antenna position significantly improves the coverage at all three heights. 

%
\subsubsection{Real-Time Camera Relocalization}
Either rotation vector (radians) $\vr\in\R^3$ or quaternion $\vq\in\R^4$ can be used to represent the camera orientation. In Table \ref{table:compare_loss} we compare the output features $\vr$ and $\vq$ when using different loss functions:  
1) {\bf W\_Euc\_Euc}: weighted sum of location Euclidean distance and quaternion Euclidean distance,
and 2) {\bf W\_Euc\_Ang}: weighted sum of location Euclidean distance and quaternion angle error. Table \ref{table:compare_loss} shows that using quaternion $\vq$ as output and including angle error in the loss function provides the best performance. The average inference time per frame is $3.87$ ms. 
\begin{table}[ht]
\vspace{-2ex}
\centering
  \caption{Comparison of loss function and output features}
  \label{table:compare_loss}
  \begin{tabular}{llll}
  \hline
  \multicolumn{1}{|l|}{}  & \multicolumn{1}{l|}{\begin{tabular}[c]{@{}l@{}}$\vo=\vr$\\W\_Euc\_Euc\end{tabular}} 
                          & \multicolumn{1}{l|}{\begin{tabular}[c]{@{}l@{}}$\vo=\vq$\\W\_Euc\_Euc \end{tabular}} 
                          & \multicolumn{1}{l|}{\begin{tabular}[c]{@{}l@{}}$\vo=\vq$\\W\_Euc\_Ang\end{tabular}} \\ 
                          \hline
  \multicolumn{1}{|l|}{Mean location error} & \multicolumn{1}{l|}{$9.2124$ cm} & \multicolumn{1}{l|}{$10.3212$ cm}   & \multicolumn{1}{l|}{\textbf{8.2937} cm}  \\ 
  \hline
  \multicolumn{1}{|l|}{Mean orientation error} & \multicolumn{1}{l|}{$8.8237^{\circ}$}  & \multicolumn{1}{l|}{$5.6327^{\circ}$}  & \multicolumn{1}{l|}{$\textbf{3.5262}^{\circ}$}\\ 
  \hline                                                                    
  \end{tabular}
\end{table}
\subsubsection{Performance of the End-to-End Solution}
\begin{figure}
     \centering
     \begin{subfigure}[b]{0.4\textwidth}
         \centering
         \includegraphics[width=\textwidth]{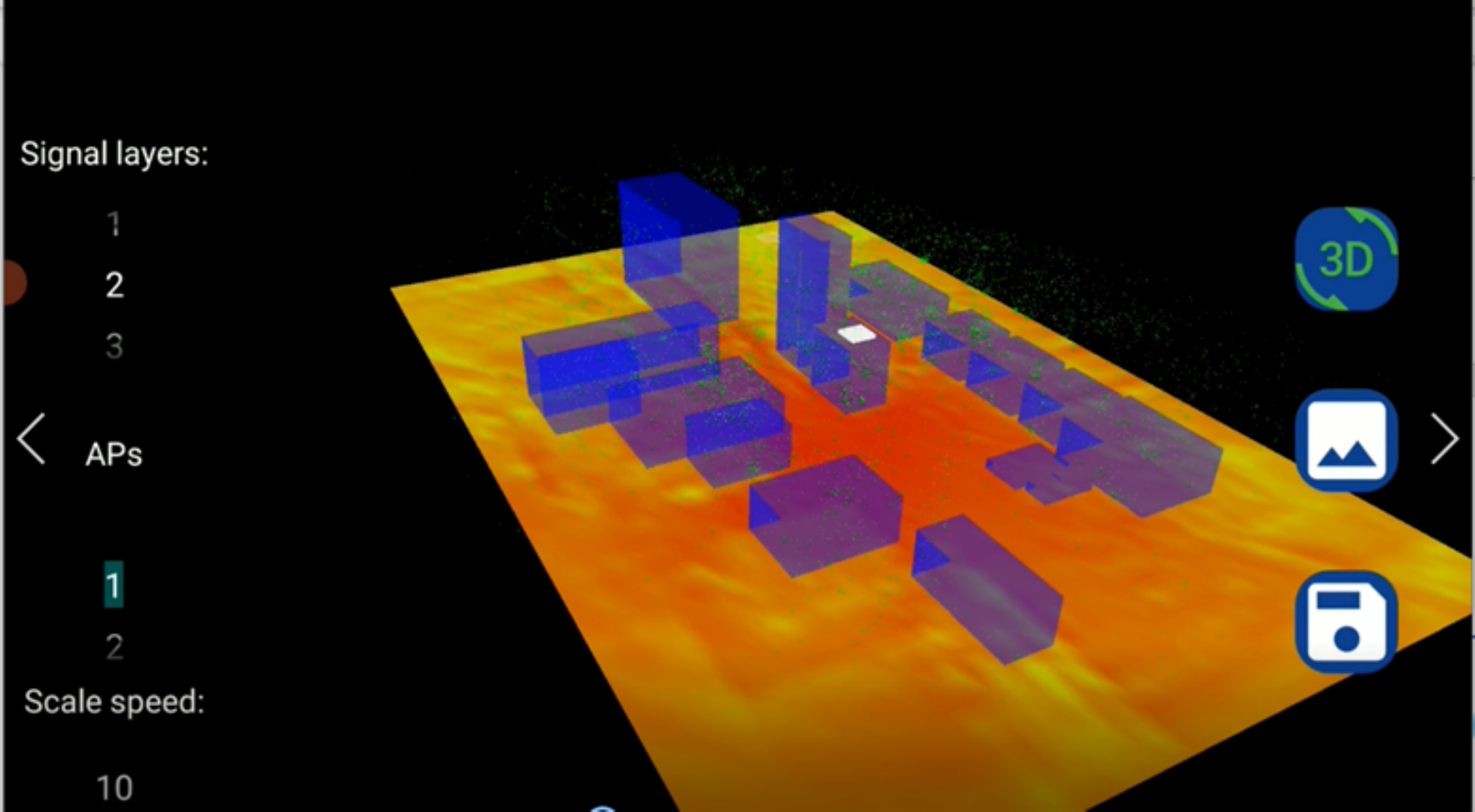}
         \caption{Interactive birdview of the reconstructed 3D virtual network. User can choose different height layers and antenna positions to switch the view of the network.}
         \label{fig:birdview}
         \vspace{-1.5ex}
     \end{subfigure}
     \begin{subfigure}[b]{0.395\textwidth}
         \centering
         \includegraphics[width=\textwidth]{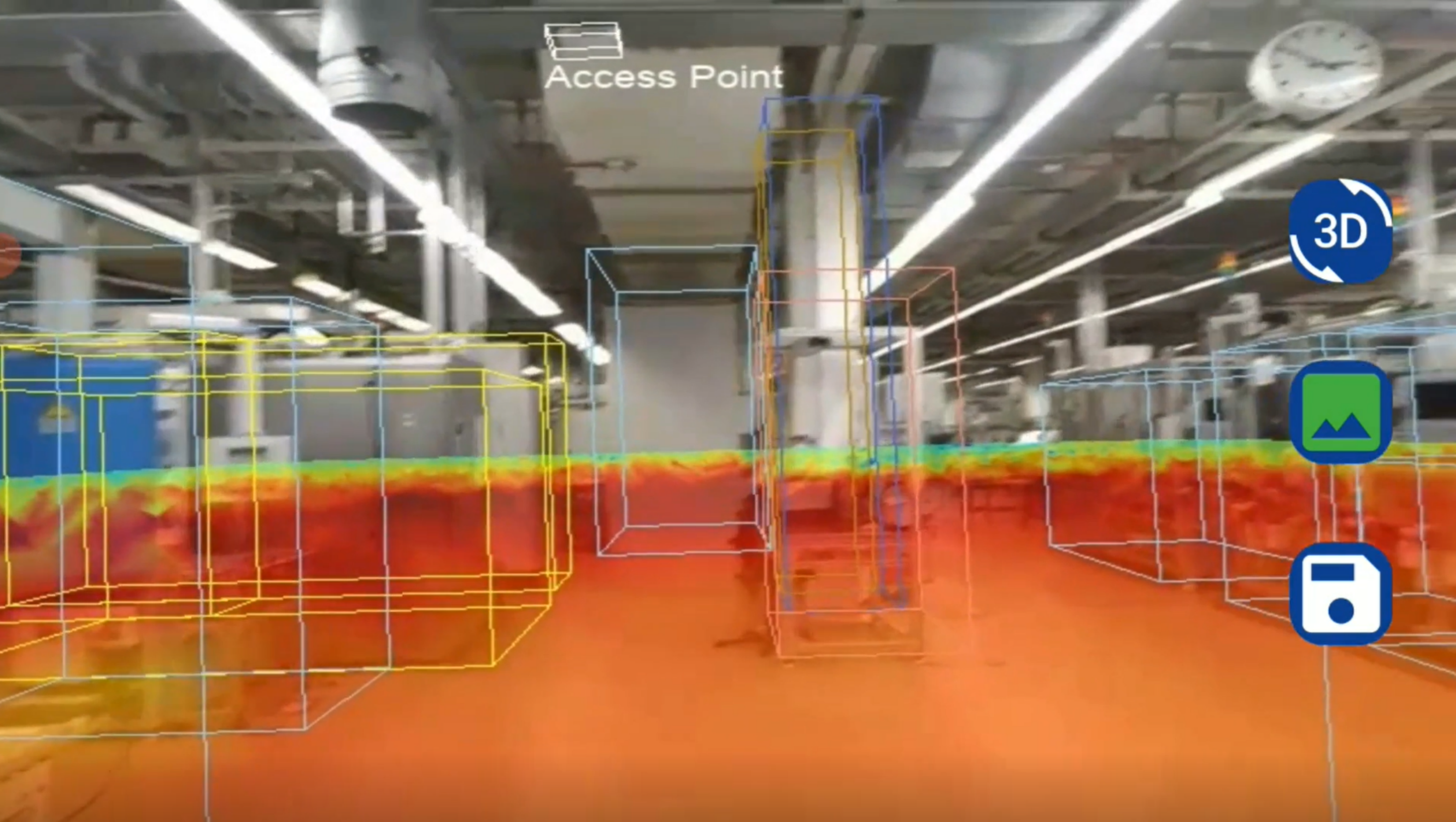}
         \caption{Real-time \ac{AR}-assisted network monitoring, with augmented radio propagation map and detected objects.}
         \label{fig:AP_view}
         \vspace{-2.5ex}
     \end{subfigure}
        \caption{Developed \ac{UI} of the Android app}
        \label{fig:Appview}
        \vspace{-5ex}
\end{figure}

For the \ac{AR}-empowered network visualization, we achieve the end-to-end latency (including data processing, data transmission, control message transmission, and deep learning-based model inference) of $31.83$ ms per frame, i.e., frame rate of over $30$ fps. We can also interact with the virtual 3D model as shown in Fig. \ref{fig:birdview} or monitor the network with an \ac{AR} view in real time as shown in Fig. \ref{fig:AP_view}.
%

\section{Conclusion and Future Work}\label{sec:concl}
We proposed a novel, interactive, and \ac{AR}-assisted framework for industrial network planning service. We provided an end-to-end solution, which receives visual and sensory data from the mobile device, reconstructs the 3D industrial network environment, performs network planning in the server, and visualizes the network with \ac{AR} on the mobile device. Several building blocks were developed: 1) \ac{SLAMORE} algorithm  to reconstruct the customized radio propagation environment for the factory, 2) ray tracing-based radio map generation and network planning, and 3) deep learning- and sensor fusion-based real-time camera relocalization to enable the \ac{AR} features. Finally, we conducted the \ac{PoC} of the proposed end-to-end solution in a Bosch factory in Germany and showed good coverage of the optimized antenna location, as well as high accuracy in both environment reconstruction and camera relocalization. We also achieved the real-time end-to-end latency of less than $32$ ms per frame for the \ac{AR}-empowered network monitoring. The future work includes further scaling up and accelerating the algorithms for larger factory space.  
%
%
\section*{Acknowledgment}
This work was supported by the German Federal Ministry of Education and Research (BMBF) project KICK [16KIS1102K]. We thank Csaba Szabo for his contribution to this project. Partial contents of this paper appear in \cite{Liao20} and \cite{HuCamLoc21}.
\acrodef{3GPP}{3rd generation partnership project}
\acrodef{3D}{three-dimensional}
\acrodef{2D}{two-dimensional}
\acrodef{5G}{the fifth generation}
\acrodef{6G}{the sixth generation}
\acrodef{AI}{artificial intelligence}
 \acrodef{ABS}{almost blank subframe}
 \acrodef{AES}{advanced encryption standard}
\acrodef{APE}{absolute percentage error}
\acrodef{AR}{augmented reality}
\acrodef{AP}{access point}
\acrodef{AGV}{automated guided vehicle}
    \acrodef{BS}{base station}
		\acrodef{BA}{bundle adjustment}
    \acrodef{CDF}{cumulative distribution function}
		\acrodef{CEVP}{constrained eigenvalue problem}
    \acrodef{CSI}{channel state information}
    \acrodef{CQI}{channel quality indicator}
		\acrodef{CNN}{convolutional neural network}
\acrodef{DoF}{degrees of freedom}
\acrodef{DL}{downlink}
\acrodef{DUDe}{downlink and uplink decoupling}
\acrodef{DNN}{deep neural network}
\acrodef{eICIC}{enhanced intercell interference coordination}
\acrodef{ESD}{energy spectral density}
\acrodef{ECDSA}{elliptic curve digital signature algorithm}
\acrodef{FDD}{frequency division duplex}
    \acrodef{FDMA}{frequency division multiple access}
		\acrodef{fps}{frame per second}
   \acrodef{GP}{Gaussian process}
    \acrodef{GPS}{global positioning system}
		\acrodef{GUI}{graphical user interface}
\acrodef{HetNet}{heterogeneous network}
\acrodef{HOG}{histogram of oriented gradients}
    \acrodef{ICI}{inter-cell interference}
		\acrodef{IMI}{inter-mode interference}
		\acrodef{IMU}{inertial measurement unit}
		\acrodef{IoT}{internet of things}
\acrodef{KL}{Kullback-Leibler}
\acrodef{LTE}{long term evolution}

\acrodef{mAP}{mean average precision}
\acrodef{MAC}{media access control}
\acrodef{MAPE}{mean absolute percentage error}
\acrodef{MIMO}{multiple-input and multiple-output}
\acrodef{MRU}{minimum resource unit}
\acrodef{MLP}{multi-layer perceptron}
\acrodef{mmWave}{millimeter wave}
\acrodef{NLES}{nonlinear equation system}
   \acrodef{OFDM}{orthogonal frequency division multiplexing}
	\acrodef{ORB}{oriented FAST and rotated BRIEF}
    \acrodef{PDF}{probability density function}
    \acrodef{PHY}{physical layer}
		\acrodef{PSD}{power spectral density}
    \acrodef{PRB}{physical resource block}
    \acrodef{PoC}{proof-of-concept}
   \acrodef{QoE}{quality of experience}
    \acrodef{QoS}{quality of service}
    \acrodef{RAN}{radio access network}
		\acrodef{RB}{resource block}
		\acrodef{RBS}{removal of bottleneck services}
		\acrodef{R-FCN}{region-based fully convolutional networks}
		\acrodef{RMDI}{resource muting for dominant interferer}
		\acrodef{RMSE}{root mean square error}
		\acrodef{ROI}{region of interest}
		\acrodef{RPN}{region proposal network}
    \acrodef{RRM}{radio resource management}
    \acrodef{RTP}{real-time transport protocol}
    \acrodef{RSRP}{reference signal received power}
		\acrodef{RU}{resource unit}
		\acrodef{RX}{receiver}
 \acrodef{SAFP}{successive approximation of fixed point}
 \acrodef{SE}{secure element}
    \acrodef{SDN}{software defined network}
     \acrodef{SDK}{software development kit}
    \acrodef{SNR}{signal-to-noise ratio}
    \acrodef{SINR}{signal-to-interference-plus-noise ratio}
\acrodef{SIR}{signal-to-interference ratio}
\acrodef{SIF}{standard interference function}
\acrodef{SLAM}{simultaneous localization and mapping}
\acrodef{SLAMORE}{simultaneous localization and mapping with object recognition}
\acrodef{SoC}{system-on-chip}
\acrodef{SSD}{single-shot multibox detector}
\acrodef{SPI}{serial peripheral interface}
    \acrodef{SVM}{support vector machine}
		\acrodef{SVD}{singular value decomposition}
    \acrodef{TCP}{transmission control protocol}
		\acrodef{TLS}{transport layer security}
		\acrodef{TDD}{time division duplex}
    \acrodef{TDMA}{time division multiple access}
		\acrodef{TTI}{transmission time interval}
		\acrodef{TX}{transmitter}
		\acrodef{UART}{universal asynchronous receiver-transmitter}
		\acrodef{UE}{user equipment}
		\acrodef{UI}{user interface}
		\acrodef{UL}{uplink}
		\acrodef{UAV}{unmanned aerial vehicle}
    \acrodef{WLAN}{wireless local area network}
\acrodef{YOLO}{you only look once}
\bibliographystyle{IEEEtran}
\bibliography{refs}
\end{document}